# Block Network Error Control Codes and Syndrome-based Complete Maximum Likelihood Decoding

Hossein Bahramgiri and Farshad Lahouti
Wireless Multimedia Communications Laboratory
School of ECE, College of Engineering
University of Tehran

*Abstract-* In this paper, network error control coding is studied for robust and efficient multicast in a directed acyclic network with imperfect links. The block network error control coding framework, BNEC, is presented and the capability of the scheme to correct a mixture of symbol errors and packet erasures and to detect symbol errors is studied. The idea of syndrome-based decoding and error detection is introduced for BNEC, which removes the effect of input data and hence decreases the complexity. Next, an efficient three-stage syndrome-based BNEC decoding scheme for network error correction is proposed, in which prior to finding the error values, the position of the edge errors are identified based on the error spaces at the receivers. In addition to bounded-distance decoding schemes for error correction up to the refined Singleton bound, a complete decoding scheme for BNEC is also introduced. Specifically, it is shown that using the proposed syndrome-based complete decoding, a network error correcting code with redundancy order $\delta_t$ for receiver $t$, can correct $\delta_t$-1 random additive errors with a probability sufficiently close to 1, if the field size is sufficiently large. Also, a complete maximum likelihood decoding scheme for BNEC is proposed. As the probability of error in different network edges is not equal in general, and given the equivalency of certain edge errors within the network at a particular receiver, the number of edge errors, assessed in the refined Singleton bound, is not a sufficient statistic for ML decoding.

*Index terms*- Network coding, network error control coding, multicast, bounded-distance and maximum likelihood decoding, Singleton bound, syndrome-based decoding and error detection.

I. INTRODUCTION

The joint design of network and channel codes is an enabling technique for robust and efficient communications over networks with imperfect links. In this paradigm, this article deals with the design of block network error control codes and syndrome-based error correction and detection in a directed acyclic network. More specifically, we consider the scenario in which the network is subject to a mixture of additive random symbol errors and packet erasures with possibly different probabilities in different links.

*A) Network Coding for Multicast*
Ahlswede *et al*. introduce a new target for multicast in a directed acyclic error-free network by evaluating the maximum multicast rate (often referred to as multicast capacity) and showing that it can be achieved by network coding [1]. Next, Li *et al*. show that the capacity can be achieved even by linear coding and finite alphabet size [2]. Koetter and Medard present an algebraic scheme for network coding and studied information flows [3]. Ho *et al*. propose a simple random decentralized scheme for network coding [4], in which the probability of success is enhanced by enlarging the Galois field size. Jaggi *et al*. in [5] introduce a centralized algorithm to find the coding solution in polynomial time. They introduced deterministic linear network coding (DLIF), a completely deterministic algorithm, and random linear network coding (RLIF), which is random in the middle stages. Fragouli *et al*. in [6], develop a distributed and deterministic method for network code design by translating the problem to a graph coloring problem. More recently, Jabbarihagh and Lahouti presented a decentralized approach based on learning to design network codes, which may be viewed as a smart random search [7].

*B) Robustness via Network Coding*
Robustness in network coding is studied from two different perspectives. In the first category, link failure is considered. Koetter and Medard in [3], show that there is a network code which can support a determined rate, $k$, for all failure patterns that do not reduce the multicast capacity below the set value. In [5], Jaggi *et al*. introduce a network code for this case. Other works in this category include [8]-[13]. In [12], a low complexity code, referred to as robust network coding, RNC($h,k$) is proposed for a network with failure-free multicast capacity of $h$, in which a trade off is set between the transmission rate, $k$, and robustness.
The second approach is started by the work of Cai and Yeung [14], in which network error correcting codes are introduced. The target is to correct a determined number of edge errors. The work is explained in detail in [15], [16]. The Hamming, Singleton and Gilbert-Varshamov bounds for network codes are studied. Zhang in [17], and Yang *et al*. in [18], independently present a refined Singleton bound, which depends on each receiver min-cut value, and thus, may be different for receivers. Minimum rank for linear network codes is introduced in [17], which plays the role of the minimum distance in classic error correcting codes. In [19], minimum distance for linear network codes is presented and based on which, the capability of a network code for error correction, error detection and erasure correction is characterized. Also, the equivalency of minimum rank and minimum distance is shown.

For construction of network error correcting codes, in [20], Jaggi *et al.* present a scheme, in which the coding is executed on the basis of a packet of symbols, which we refer to it as packet-based coding. It is further assumed that the error positions (edges which may encounter error) remain fixed over a packet duration. We refer to the new bound for this case as the *packet-based Singleton bound*. As expected and discussed in section III.D, the network code designed based on this bound, as opposed to the singleton bound, is able to correct a much smaller number of error-patterns. The scheme of [20] attains the packet-based Singleton bound with high probability. Yang *et al.* present a symbol-based coding scheme to achieve the refined Singleton bound in [18], based on the algebraic approach in [3]. With the same objective, Matsumoto in [21] suggests a network code based on the preservative approach of [5], and also studies the probability of successful random code construction. Koetter and Kschischang study error correction in random network codes [22] to achieve the packet-based refined Singleton bound. The code is designed at the transmitter, based on vector spaces which are preserved through the network, and the structure of the network does not affect the design.

For decoding, Jaggi *et al.* in [20], present an algorithm for their packet-based coding scheme, aiming at the packet-based Singleton bound. In this scheme, linear decoding equations of data and error are constructed, and using a proxy error matrix the error is estimated. Considering the number of symbols in a packet, $K$, the decoding delay and complexity grows proportional to $K$ and $K^3$, respectively. Matsumoto proposes a simple exhaustive search for decoding in the symbol level for the network code presented in [21], in which the received symbol is checked against all information symbols and correctable error symbols. In [23][24], a minimum rank decoding scheme is presented for linear network error correction codes in a multicast application. Attempting to separate the error and the input message parts for efficient decoding of the received signals, they present a method based on Gaussian elimination. They also present a fast decoding scheme that for certain error events can locate the error positions first to simplify error computation. The scheme is packet-based and assumes that (i) the error-pattern remains unchanged over a packet and, (ii) a sufficient number of errors in these positions occur within a packet. Considering the second assumption, the decoding scheme does not attain the packet-based refined Singleton bound. In [23], they present the probability of successful decoding in their packet-based setting, assessing decoding beyond the design error correction capability. Balli *et al.* study the success probability of random linear network error correction codes to be decodable in a multicast session in the presence of additive errors [25].

*C) Outline*

In this work, we study network error control coding, for a mixture of random additive errors and erasures. The system model and primary definitions are presented in section II. We set up a block network error control (BNEC) code framework, for multicast in an acyclic network in section III. The BNEC($h+,k$) takes a $k$-symbol input data, and produces a fixed-length $h_t$-symbol codeword at the receiver $t$, where $h_t$ is the min-cut of the receiver $t$. Depending on the added redundancy of order $\delta_t = h_t - k$, the receivers decoding capability is enhanced in face of errors and erasures. We then assess the BNEC error and erasure correction and error detection performance, based on the refined Singleton bound [17][18]. We refer to the results as the BNEC *bounded-distance* capabilities, using which a strategy for construction of a BNEC code is presented. A detailed code design algorithm is then proposed in section VIII.

Next, in section IV, we study the decoding and error detection of BNEC codes. Here, we introduce the syndrome-based scheme for BNEC code, which removes the effect of the input data in the process of decoding and detection, and hence substantially reduces the associated complexity and memory requirement.

To further enhance the BNEC decoding efficiency, in section V, we present a three-stage syndrome-based network decoding scheme. The approach is composed of the following stages: error detection, finding the positions of the errors, and finding the value of the errors in these positions. The presented decoding schemes, consider the decoding up to the refined Singleton bound for block network error control codes. Therefore, in line with classic channel coding, we refer to them as the BNEC bounded-distance decoding.

In section VI, we present the Complete Maximum Likelihood (ML) decoding, for block network error control codes. The scheme offers the following benefits in comparison to the bounded-distance decoding: (1) Although, the BNEC is designed for error correction up to the refined Singleton bound, but there are a large number of correctable errors beyond this bound, which is the key to the design of a *BNEC complete decoder*, (2) Unlike the classic channel coding design, in which the probabilities of error in different positions are considered the same, here, the probabilities of error in different network edges may be different, (3) There may be different sets of network edges with the same effects at a certain receiver. The proposed *BNEC ML decoder* takes the latter two points into account to find the most likely error at the receivers. The error correction capabilities of the complete decoder are studied, which leads to an interesting bound: A BNEC($h+,k$), which is designed to correct up to $\lfloor \delta_t / 2 \rfloor$ errors, can correct up to $\delta_t - 1$ random additive errors with a probability approaching 1, as the field size is increased.

In section VII, we compare the performance of the proposed BNEC bounded-distance and complete ML decoding schemes. Also, we analyze their complexity and memory requirement, employing the two syndrome-based approaches. In section VIII, an algorithm to design a BNEC code is presented. Finally, the paper is concluded in section IX.

## II. PRELIMINARIES

*A) Network Model*
Consider a directed acyclic network, represented by the graph $G=(V,E)$, where $V$ and $E$ denote the set of nodes and the set of edges of the graph, respectively. For the edge $l=(a,b)$, $a=start(l)$ and $b=end(l)$. The sequence of edges, $f^t = [(x,b_1),(a_2,b_2),...,(a_{m-1},b_{m-1}),(a_m,y)]$, is a path from the node $x$ to the node $y$, if for $i=2,...,m$, $b_{i-1} = a_i$. We focus on a multicast transmission, with a single transmitting node, denoted by $s$, and a set of receivers with the same data requests, denoted by $T$. We assume synchronous transmission and unit transmission rate for each edge. We can model an edge with a positive integer rate $r$, by $r$ parallel edges with rate one. We define $E^a$ as the set of *active edges* in the network, *i.e.*, the set of edges which transmit data in a multicast session. We assign an index to each active edge $l_i$, denoted by $n(l_i)=i$, such that if there is a path from $end(l_i)$ to $start(l_j)$, then $j>i$. This numbering scheme is possible in an acyclic network. Also, we define the set of *active edges of the receiver t* as $E_t$, as the set of edges $l \in E^a$, for which there is at least one path from $end(l)$ to the receiver $t$. The set of all active edges $l'$, such that, $end(l')=start(l)$, is denoted by $In(l)$.
The min-cut value of each receiver $t$, is denoted by $h_t$, and also $h=\min\{h_t\}$ is the error-free multicast capacity of the network [1], which is shown to be attained by linear codes in [2].

*B) Noise Model*
Each active edge of the network transmits a symbol from the finite field $F_q$ in each time unit, and may encounter erasure or error, here referred to as noise. The channel model of the edge $l$ is a cascade of erasure and error, which is generally assumed to be independent in practical networks, as depicted in Figure 1, for $q=3$. The probabilities of error and erasure at the edge $l$ are denoted by $p_{err}(l)$ and $p_{ers}(l)$, respectively. The additive error is assumed to have a uniform distribution over $F_q$, for which we have

$$p_{ij}(l) = \begin{cases} 1 - p_{err}(l) & i = j \\ p_{err}(l)/(q-1) & i \neq j \end{cases}. \quad (1)$$

in which $p_{ij}(l)$ is the transition probability from symbol value $i$ to $j$ ($0 \leq i,j \leq q-1$), for the edge $l$. In the event of an erasure in an edge, the edge output is considered to be zero, while its position is known in the subsequent edges. Channel models with a mixture of errors and erasures are motivated for communications over wireless networks or a mixture of wireless and wireline networks [26]-[28].
The following definitions are used throughout this paper (Note that we only consider the set of active edges and remove the other edges):
1- *Noise-vector*: We define the noise-vector, **e**, as a vector of size $|E^a| \times 1$, in which the *i*'th element indicates the value of noise in the *i*'th edge of the network.

2- *Noise-pattern*: A noise-pattern $\Phi=\{j_1,...j_b\}^{\tau}$ ($\tau$ denotes the transpose operation) is a set comprising the indices of the edges, in which the noise may be present.
The number of noisy edges in a network is the hamming-weight of the vector **e**, $hw(\mathbf{e})$. The number of noisy edges for the receiver $t$, is the number of noisy edges, in the edge set $E_t$, which is less than or at most equal to $hw(\mathbf{e})$.
Note that, we similarly define error-vector and error-pattern, considering only errors, and also define erasure-vector and erasure-pattern, considering only erasures.
3- *N(A,B)*: Considering $A = (\alpha_1,\cdots,\alpha_{|T|})$ and $B = (\beta_1,\cdots,\beta_{|T|})$, we define $N(A,B)$ as the set of all noise-vectors, in which the number of erasures and errors are equal to or less than $\alpha_t$ and $\beta_t$, respectively, $\forall t \in T$.
4- *N(Γ)*: Considering $\Gamma = (\gamma_1,...,\gamma_{|T|})$, we define $N(\Gamma)$ as the set of all error-vectors, in which the number of errors in $E_t$ is equal to or less than $\gamma_t$, $\forall t \in T$.
5- *Uniform Network*: We define a uniform network as a network for which all the edges have the same error probabilities and the same erasure probabilities.

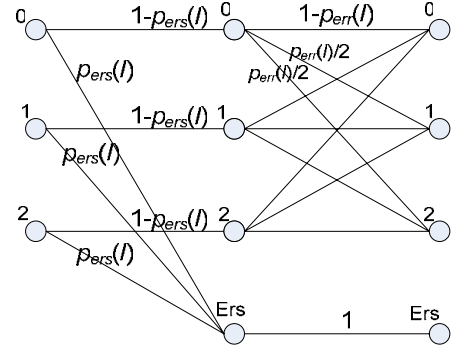

Figure 1- The channel model for the edge $l$, which is a cascade of error and erasure ($q = 3$).

## III. BLOCK NETWORK ERROR CONTROL CODES

*A) BNEC Framework*
Here, we present the block network error control coding framework, for multicast transmission in a network (see Figure 2). The BNEC($h+,k$) scheme, is a joint network-channel code, facilitating a multicast rate $k$, in presence of error and erasure on the edges of a network with multicast capacity $h$, and unicast capacity $h_t \geq h$, $\forall t \in T$.

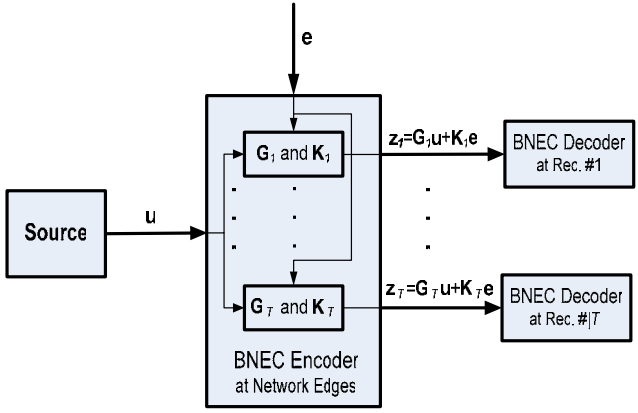

Figure 2- The framework of BNEC coder and decoder

The input data is assumed to be i.i.d. symbols from the finite field $F_q$, of size $q$. In one network use (one time unit), $k$ input symbols, $\mathbf{u}=[u_1,\ldots,u_k]^\tau$ $0\leq u_i \leq q-1$, $1\leq i \leq k$, are sent to the receivers. We next consider several key definitions.

1- *Global Encoding Vector*: Each edge $l_i$, transmits a symbol, $y_i$, as a linear combination of the input symbols and the noise values of the network edges,

$$y_i = \sum_{j=1}^{k} g_{i,j} u_j + \sum_{j=1}^{i} \kappa_{i,j} e_j$$
$$= \mathbf{gev}_i^\tau \cdot \begin{bmatrix} \mathbf{u} \\ \mathbf{e} \end{bmatrix} \quad (2)$$

where, the vector $\mathbf{gev}_i = [g_{i,1},\ldots,g_{i,k},\kappa_{i,1},\ldots,\kappa_{i,|E^a|}]^\tau$ is the global encoding vector of the edge $l_i$, for which $\kappa_{i,j}=0$ $j>i$, as the network graph is acyclic.

2- *Local Encoding Vector*: The local encoding vector of the edge $l_i$, $\mathbf{m}_i$, is a vector of size $|In(l_i)|$, comprising the combination coefficients of the edges in the set $In(l_i)$, to form $\mathbf{gev}_i$. The element of $\mathbf{m}_i$ corresponding to the edge $l \in In(l_i)$ is denoted by $m_i(l)$. We have

$$\mathbf{gev}_i = \sum_{l \in In(l_i)} m_i(l) \mathbf{gev}_{n(l)} + e_i \cdot \mathbf{1}_{k+|E^a|,k+i}, \quad (3)$$

in which the vector $\mathbf{1}_{a,b}$ is a vector of size $a \times 1$, with a 1 in position $b$, and zero elsewhere. Also, $n(l)$ is the index of the edge $l$.

3- *Received Vector*: Each receiver $t \in T$, receives $h_t$ symbols from its $h_t$ input edges. We define $\mathbf{z}_t(\mathbf{u},\mathbf{e})$ as the received vector of the receiver $t$, of size $h_t \times 1$, which can be written as follows,

$$\mathbf{z}_t(\mathbf{u},\mathbf{e}) = \mathbf{G}_t \mathbf{u} + \mathbf{K}_t \mathbf{e} = [\mathbf{G}_t | \mathbf{K}_t] \cdot \begin{bmatrix} \mathbf{u} \\ \mathbf{e} \end{bmatrix}. \quad (4)$$

The rows of the matrix $[\mathbf{G}_t | \mathbf{K}_t]$ are the transposed version of global encoding vectors of the input edges of the receiver $t$. The matrices $\mathbf{G}_t = [\mathbf{g}_1,\cdots,\mathbf{g}_k]$, of size $h_t \times k$ and $\mathbf{K}_t = [\mathbf{k}_1,\cdots,\mathbf{k}_{|E^a|}]$, of size $h_t \times |E^a|$, are referred to as the *Data-Transformation* and *Noise-Transformation Matrices*. The data-transformation matrix, maps an input vector in the space $F_q^k$, to a vector $\mathbf{v}_t = \mathbf{G}_t \mathbf{u}$ in the column space of the matrix $\mathbf{G}_t$. Also, the noise-transformation matrix, maps the error-vector $\mathbf{e}$, to a vector referred to as the *coded error-vector*, $\mathbf{c}_t = \mathbf{K}_t \mathbf{e}$, in the column space of the matrix $\mathbf{K}_t$.

Generally speaking, a BNEC code is designed in two stages: 1- Selection of the set of active edges, $E^a$, such that the min-cut of each receiver $t$, $h_t$, is accommodated. 2- Obtaining the local encoding vector for each active edge. This results in the data and noise transformation matrices for all the receivers. A general design strategy is presented in section III.C, and a specific design algorithm is elaborated in section VIII.

*B) Error Detection & Bounded-Distance Decoding Capabilities*

Before designing a BNEC code, we study its error control capabilities in presence of both errors and erasures, while bounded-distance decoding is used for error correction. Later in section VI, the complete ML decoding for BNEC is presented which provides considerable benefits over the bounded-distance decoding.

***Theorem 1:*** If a block network error control code can correct all noise-vectors in $N(A,B)$, in which $A = (\alpha_1,\ldots,\alpha_{|T|})$ and $B = (\beta_1,\cdots,\beta_{|T|})$, then it can correct all noise-vectors in $N(A',B')$, in which $A' = (\alpha'_1,\cdots,\alpha'_{|T|})$ and $B' = (\beta'_1,\cdots,\beta'_{|T|})$, if

$$\alpha'_t + 2\beta'_t \leq \alpha_t + 2\beta_t, \forall t \in T. \quad (5)$$

*Proof*- We assume that there are two different input vectors, $\mathbf{u}_1$ and $\mathbf{u}_2$, with two noise-vectors, $\mathbf{e}_1$ and $\mathbf{e}_2$, with at most $\alpha'_t$ erasures and $\beta'_t$ errors in $E_t$, for all $t$, satisfying the Equation (5), such that the received vectors are the same for both cases at a receiver $t$: $\mathbf{z}_t(\mathbf{u}_1,\mathbf{e}_1) = \mathbf{z}_t(\mathbf{u}_2,\mathbf{e}_2)$. We have,

$$\mathbf{G}_t \mathbf{u}_1 + \mathbf{K}_t \mathbf{e}_1 = \mathbf{G}_t \mathbf{u}_2 + \mathbf{K}_t \mathbf{e}_2, \quad (6)$$

that can be rewritten as follows,

$$\mathbf{G}_t \mathbf{u}_1 + \mathbf{K}_t (\mathbf{e}_1^{\alpha'_t} + \mathbf{e}_1^{\beta'_t}) = \mathbf{G}_t \mathbf{u}_2 + \mathbf{K}_t (\mathbf{e}_2^{\alpha'_t} + \mathbf{e}_2^{\beta'_t}) \quad (7)$$

where, $\mathbf{e}_i$, $i=1,2$, is decomposed to $\mathbf{e}_i^{\alpha'_t}$ with only erasures and $\mathbf{e}_i^{\beta'_t}$ with only errors. Note that, the superscript shows the maximum possible hamming-weight of the noise-vector. The two error-vectors $\mathbf{e}_1^{\alpha'_t}$ and $\mathbf{e}_2^{\alpha'_t}$ have non-zero values at the same positions. Thus,

$$\mathbf{G}_t.\mathbf{u}_1 + \mathbf{K}_t.(\mathbf{e}_3^{\alpha'_t} + \mathbf{e}_1^{\beta'_t}) = \mathbf{G}_t.\mathbf{u}_2 + \mathbf{K}_t \mathbf{e}_2^{\beta'_t}, \quad (8)$$

where, $\mathbf{e}_3^{\alpha'_t} = \mathbf{e}_1^{\alpha'_t} - \mathbf{e}_2^{\alpha'_t}$.

*Case 1)* $\alpha'_t - \alpha_t \geq 0$: From Equation (5), we have $\alpha'_t - \alpha_t \leq 2(\beta_t - \beta'_t) = 2\sigma_t$. We can decompose $\mathbf{e}_3^{\alpha'_t}$ as $\mathbf{e}_3^{\alpha'_t} = \mathbf{e}_3^{\alpha_t} + \mathbf{e}_3^{\sigma_t} + \mathbf{e}_4^{\sigma_t}$. Then, Equation (8) can be rewritten as follows,

$$\mathbf{G}_t.\mathbf{u}_1 + \mathbf{K}_t.(\mathbf{e}_3^{\alpha_t} + \mathbf{e}_4^{\beta_t}) = \mathbf{G}_t.\mathbf{u}_2 + \mathbf{K}_t\mathbf{e}_2^{\beta_t} \qquad (9)$$

where, $\mathbf{e}_4^{\beta_t} = \mathbf{e}_3^{\sigma_t} + \mathbf{e}_1^{\beta'_t}$ and $\mathbf{e}_2^{\beta_t} = -\mathbf{e}_4^{\sigma_t} + \mathbf{e}_2^{\beta'_t}$. The Equation (9) shows that two noise-vectors with at most $\alpha_t$ erasures and at most $\beta_t$ errors can not be corrected at receiver $t$ for the inputs $\mathbf{u}_1$ and $\mathbf{u}_2$, which is a contradiction.

*Case 2)* $\alpha_t - \alpha'_t > 0$: From Equation (5), we have $2\sigma_t = 2(\beta'_t - \beta_t) \leq \alpha_t - \alpha'_t$. We can decompose $\mathbf{e}_i^{\beta'_t}$ as $\mathbf{e}_i^{\beta'_t} = \mathbf{e}_i^{\beta_t} + \mathbf{e}_i^{\sigma_t}$, for $i=1,2$. Thus, Equation (8) can be rewritten as follows,

$$\mathbf{G}_t.\mathbf{u}_1 + \mathbf{K}_t.(\mathbf{e}_3^{\alpha_t} + \mathbf{e}_2^{\beta_t}) = \mathbf{G}_t.\mathbf{u}_2 + \mathbf{K}_t\mathbf{e}_2^{\beta_t}, \qquad (10)$$

where $\mathbf{e}_3^{\alpha_t} = \mathbf{e}_3^{\alpha'_t} + \mathbf{e}_1^{\sigma_t} - \mathbf{e}_2^{\sigma_t}$. The Equation (10) shows that two noise-vectors with at most $\alpha_t$ erasures and at most $\beta_t$ errors can not be corrected at the receiver $t$ for the inputs $\mathbf{u}_1$ and $\mathbf{u}_2$, which is a contradiction. □

Note that, in Theorem 1, if we consider $\alpha'_t = \beta_t = 0$, $\forall t \in T$, then we have the following result: If a BNEC code can correct $\alpha_t$ erasures in $E_t$, $\forall t \in T$, then it can correct $\lfloor \alpha_t/2 \rfloor$ errors in $E_t$, $\forall t \in T$. Also, if we consider $\alpha_t = \beta'_t = 0$, $\forall t \in T$, then we have the following result: If the BNEC can correct $\beta_t$ errors in $E_t$, $\forall t \in T$, then it can correct $2\beta_t$ erasures in $E_t$, $\forall t \in T$. These two results are presented in [19] based on the Singleton bound, and in [17][18] based on the refined Singleton bound.
The relation between error detection and correction capabilities of a network error control code is presented in [19]. Here, we present it in the BNEC framework for the mixture of erasures and errors.

**Theorem 2:** A block network error control code can correct all noise-vectors in $N$(A,B), in which $A = (\alpha_1,...,\alpha_{|T|})$ and $B = (\beta_t,\cdots,\beta_{|T|})$, if and only if it can detect all error-vectors in $N(\Gamma)$, in which $\Gamma = (\alpha_1 + 2\beta_1,...,\alpha_{|T|} + 2\beta_{|T|})$.

*Proof-Direct-* Assume the code can not correct two noise-vectors, $\mathbf{e}_1 = \mathbf{e}_1^{\alpha_t} + \mathbf{e}_1^{\beta_t}$ and $\mathbf{e}_2 = \mathbf{e}_2^{\alpha_t} + \mathbf{e}_2^{\beta_t}$, added to two different input vectors $\mathbf{u}_1$ and $\mathbf{u}_2$, in which $\mathbf{e}_i^{\alpha_t}$ and $\mathbf{e}_i^{\beta_t}$ are the erasure and error parts of $\mathbf{e}_i$, $i=1,2$. Then, we have

$$\mathbf{G}_t.\mathbf{u}_1 + \mathbf{K}_t.(\mathbf{e}_1^{\alpha_t} + \mathbf{e}_1^{\beta_t}) = \mathbf{G}_t.\mathbf{u}_2 + \mathbf{K}_t.(\mathbf{e}_2^{\alpha_t} + \mathbf{e}_2^{\beta_t})$$
$$\Rightarrow \mathbf{G}_t.\mathbf{u}_1 + \mathbf{K}_t.(\mathbf{e}_3^{\alpha_t} + \mathbf{e}_3^{2\beta_t}) = \mathbf{G}_t.\mathbf{u}_2, \qquad (11)$$

where, $\mathbf{e}_3^{\alpha_t} = \mathbf{e}_1^{\alpha_t} - \mathbf{e}_2^{\alpha_t}$, $\mathbf{e}_3^{2\beta_t} = \mathbf{e}_1^{\beta_t} - \mathbf{e}_2^{\beta_t}$. The Equation (11) shows that the receiver can not detect an error-vector with a hamming-weight of at most $\alpha_t + 2\beta_t$, which is a contradiction.

*Proof-Reverse-* Assume that the code can not detect an error-vector $\mathbf{e}_1$ in $\alpha_t + 2\beta_t$ positions. The vector can be decomposed as $\mathbf{e}_1 = \mathbf{e}_2^{\alpha_t} + \mathbf{e}_3^{\beta_t} + \mathbf{e}_4^{\beta_t}$ with their nonzero values at distinct positions. Also, $\mathbf{e}_2^{\alpha_t}$ can be written as $\mathbf{e}_2^{\alpha_t} = \mathbf{e}_3^{\alpha_t} + \mathbf{e}_4^{\alpha_t}$. Then, we have,

$$\mathbf{G}_t.\mathbf{u}_1 + \mathbf{K}_t.\mathbf{e}_1 = \mathbf{G}_t.\mathbf{u}_2$$
$$\Rightarrow \mathbf{G}_t.\mathbf{u}_1 + \mathbf{K}_t.(\mathbf{e}_3^{\alpha_t} + \mathbf{e}_4^{\alpha_t} + \mathbf{e}_3^{\beta_t} + \mathbf{e}_4^{\beta_t}) = \mathbf{G}_t.\mathbf{u}_2 \qquad (12)$$
$$\Rightarrow \mathbf{G}_t.\mathbf{u}_1 + \mathbf{K}_t.(\mathbf{e}_3^{\alpha_t} + \mathbf{e}_3^{\beta_t}) = \mathbf{G}_t.\mathbf{u}_2 - \mathbf{K}_t.(\mathbf{e}_4^{\alpha_t} + \mathbf{e}_4^{\beta_t})$$

This equation shows that the receiver can not distinguish between two noise-vectors with $\alpha_t$ erasures and $\beta_t$ errors, which is a contradiction. □

Note, if we consider $\alpha_t = 0$, then we have the following result: A block network error control code can correct all error-vectors with $\beta_t$ errors in $E_t$, $\forall t \in T$, if and only if it can detect all error-vectors with $2\beta_t$ errors in $E_t$, $\forall t \in T$. This result is presented in [19] based on the Singleton bound, and in [18] based on the refined Singleton bound.
After analyzing the relation between error correction and detection, we next study the correction capabilities of a BNEC code, based on the refined Singleton bound.

**Theorem 3:** If a block network error control code can correct all noise-vectors in $N$(A,B), then

$$\alpha_t + 2\beta_t \leq h_t - k, \forall t \in T. \qquad (13)$$

*Proof-* From theorem 1, we can deduce that the code can correct all erasure-patterns with at most $\alpha_t + 2\beta_t$ erasures in $E_t$, $\forall t \in T$. Thus, the Equation (13) can be obtained from the refined Singleton bound [17][18]. □

Note that, a BNEC code can correct simultaneously all noise-vectors in all $N$(A,B) sets, when $\alpha_t + 2\beta_t \leq h_t - k$, $\forall t \in T$.
Recently Zhang in [24] studies the capability of network code error and erasure correction in a packet-based setting, where the global encoding vectors (extended global encoding vectors in [24]) are sent in each packet header and used for decoding at the receivers. As elaborated in section III.D, in the BNEC framework, such a header may be avoided even in the presence of erasure.

*C) Design strategy for BNEC code*
As evident in Theorems 1-3, to achieve the bounded-distance capabilities of the BNEC code for error detection and error and erasure correction, it is sufficient to design a code to correct all erasure-vectors with $\delta_t = h_t - k$ erasures in $E_t$, $\forall t \in T$. Assume an input data $\mathbf{u}$ and an erasure-vector $\mathbf{e}$, with corresponding erasure-pattern set $\Phi$, with $\delta_t$ erasures in $E_t$ for all $t$. At the receiver $t$, the received vector is

$$\begin{aligned}\mathbf{z}_t(\mathbf{u},\mathbf{e}) &= [\mathbf{G}_t, \mathbf{K}_t] \cdot \begin{bmatrix} \mathbf{u} \\ \mathbf{e} \end{bmatrix} \\ &= [\mathbf{G}_t, \mathbf{K}_t^\Phi] \cdot \begin{bmatrix} \mathbf{u} \\ \mathbf{e}^\Phi \end{bmatrix}\end{aligned} \quad (14)$$

in which, $\mathbf{K}_t^\Phi$ and $\mathbf{e}^\Phi$ contain respectively, only the columns or the elements of $\mathbf{K}_t$ and $\mathbf{e}$, that correspond to the erasure-pattern set $\Phi$ and are also present in $E_t$. The matrix $[\mathbf{G}_t, \mathbf{K}_t^\Phi]$ is of size $h_t \times h_t$, and is to be designed such that for all erasure-patterns $\Phi$, the input vector $\mathbf{u}$ is solvable from Equation (14). Note that, a sufficient condition is that the matrix is invertible. However, this is not a necessary condition, as the erasure values, $\mathbf{e}^\Phi$ are not to be identified, as long as $\mathbf{u}$ is obtained.

Using the BNEC design strategy presented here, to compute the desired matrices and design the BNEC code, in Section VIII, we present an efficient algorithm based on the preservative approach of [5].

*D) Notes on Erasure, Error and Packet Structure*
For each packet, the payload consists of $K \geq 1$ symbols. The header contains the information about the positions of erasures, which as elaborated below, provides the receivers with the network erasure-pattern. An erasure occurs at the packet level, i.e., when a packet is lost (erased), all the symbols in its payload are erased, similar to communications over practical packet networks. But naturally, the error-pattern may vary over different symbol times within a packet, considering which is necessary for achieving the refined Singleton bound. The number of manageable edge errors in each symbol time, and the number of manageable erasures in each packet time, is limited by the refined Singleton bound for error detection and error and erasure correction.

As explained, a BNEC($h+,k$) code can correct up to $\delta_t = h_t - k$ erasures. To inform the receivers of the position of erasures, we consider $\delta_t$ positions in the header of the packets indicating the indices of the edges with erasure. Thus, the scheme adds $\delta_t \cdot \log_2 |E'|$ bits ($|E'| \in O(|E^a|)$) to the packet header. Note that, by the erasure of a packet, the edge indices corresponding to some previous erasures that are present in the packet header, are also lost; This, however, can be shown to be resolvable for erasure correction.

In [24], a different packet structure is proposed where (i) the global encoding vector of the edge is placed in the packet header and (ii) the error-pattern is assumed constant within a packet. Due to possible erasures within the network, the receiver then obtains different global encoding vectors from its input edges, and hence attempts to reconstructs the input symbols. This is of course without explicit knowledge of the erasure positions. Note that, the global encoding vector of an edge is of size $(k + |E'|) \cdot \log_2 q$ bits, which is much greater than the packet header overhead in the proposed scheme. Moreover, if the error-pattern is assumed constant within a packet, the coding scheme can only achieve the packet-based refined Singleton bound. In other words, by this bound, the number of edges which may encounter error within a packet duration is limited by $\delta_t$. This is in contrast to the proposed scheme, where $\delta_t$ errors may be corrected within a symbol duration, hence achieving the refined singleton bound.

Next, we study the decoding and detection schemes for a BNEC code in presence of errors and erasures in network edges. Of course, the complexity and memory requirements of a decoding scheme are important practical considerations. In the next section, we present an effective syndrome-based approach.

## IV. SYNDROME-BASED ERROR DETECTION AND DECODING

Using a BNEC code, at the receiver $t$, the received vectors for all correctable noise-vectors are distinguishable. Thus, an approach for decoding of BNEC code is the exhaustive search, similar to the scheme in [21]. In this case, the receiver can store a look-up table of all correctable received vectors and the corresponding input vectors. However, the structure of the network code may be exploited to devise more efficient decoding and detection schemes. Here, the syndrome-based decoding and detection are introduced for a BNEC code, which eliminating the effect of the input vector, substantially reduce the size of the look-up table required for decoding.

The *code space* of a BNEC code, $S_t^C$, is defined as the space comprising all the received vectors in a noise free setting: $\mathbf{z}_t(\mathbf{u},\mathbf{0}) = \mathbf{G}_t \mathbf{u} \quad \forall \mathbf{u} \in F_q^k$.

**Theorem 4**: In a BNEC code, the data-transformation matrix is full rank.

*Proof*- In a network using BNEC($h+, k$), we assume that the received vector at the receiver $t$ is noise-free: $\mathbf{z}_t(\mathbf{u},\mathbf{0})=\mathbf{G}_t\mathbf{u}$. The $k$ input data are to be decodable from $\mathbf{z}_t$, thus the rank of the matrix $\mathbf{G}_t$ is to be at least $k$. Therefore, the matrix is full rank. Note that, in [23], a similar condition is presented for a *regular* network error correcting code. □

In the matrix $\mathbf{G}_t = [\mathbf{g}_1, \ldots, \mathbf{g}_k]$, the vectors $\mathbf{g}_i$, $1 \leq i \leq k$, are linearly independent. Therefore, the code space at receiver $t$, which is equal to the column space of $\mathbf{G}_t$, is a subspace of dimension $k$ of $F_q^{h_t}$. Similar to the classic channel block coding, we define the *dual space* of a BNEC code at the receiver $t$, with a rank of $\delta_t = h_t - k$. We construct the *parity check matrix* $\mathbf{H}_t$, of size $h_t \times \delta_t$, from $\delta_t$ basis vectors of the dual space. Then, we have,

$$\mathbf{H}_t^\tau \cdot \mathbf{G}_t = \mathbf{0}. \quad (15)$$

The receiver $t$, employs the parity check matrix to check the presence of an error in the received symbol vector. The receiver $t$ computes the following vector:

$$\mathbf{s}_t(\mathbf{e}) = \mathbf{H}_t^\tau \cdot \mathbf{z}_t(\mathbf{u},\mathbf{e}), \quad (16)$$

which is a $\delta_t \times 1$ vector, referred to as the *syndrome* of the received vector $\mathbf{z}_t$. A non-zero syndrome indicates the presence of an error in the received vector:

$$\mathbf{s}_t(\mathbf{e}) = \mathbf{H}_t^\tau \mathbf{K}_t \mathbf{e}. \quad (17)$$

This equation is the direct result of Equations (4), (15) and (16), which is independent of the input vector. We now show the sufficiency of syndrome for error detection and bounded-distance decoding.

***Theorem 5:*** In BNEC($h+$, $k$), the syndrome of the received vector at a receiver is sufficient for error detection.

*Proof-* It is enough to show that the syndrome of any detectable noise-vector is nonzero. Assume $\mathbf{s}_t(\mathbf{e}) = \mathbf{0}$, for a noise-vector $\mathbf{e}$, for which $\mathbf{K}_t \mathbf{e} \neq \mathbf{0}$. Thus, $\mathbf{H}_t^\tau \mathbf{K}_t \mathbf{e} = \mathbf{0}$, which implies that $\mathbf{K}_t \cdot \mathbf{e}$ is a codeword. Therefore, for each $\mathbf{G}_t \mathbf{u}_1 \in S_t^C$, there is a vector $\mathbf{G}_t \mathbf{u}_2 \in S_t^C$, $\mathbf{u}_1 \neq \mathbf{u}_2$, such that $\mathbf{G}_t \mathbf{u}_1 + \mathbf{K}_t \mathbf{e} = \mathbf{G}_t \mathbf{u}_2$, which is a contradiction, since $\mathbf{e}$ is detectable. □

***Theorem 6:*** In BNEC($h+$, $k$), the syndrome of the received vector at a receiver is sufficient for bounded-distance decoding.

*Proof-* Assume all noise-vectors with $\alpha_t$ erasures and $\beta_t$ errors in $E_t$, $\forall t \in T$, are correctable (for example, by exhaustive search). It is enough to show that the syndrome vectors of two correctable noise-vectors, $\mathbf{e}_1$ and $\mathbf{e}_2$, with common erasure-patterns and different encoded noise-vectors, $\mathbf{c}_1 = \mathbf{K}_t \mathbf{e}_1$ and $\mathbf{c}_2 = \mathbf{K}_t \mathbf{e}_2$, are distinct. Assuming $\mathbf{s}_t(\mathbf{e}_1) = \mathbf{s}_t(\mathbf{e}_2)$, we have

$$\mathbf{H}_t^\tau \mathbf{K}_t (\mathbf{e}_1 - \mathbf{e}_2) = \mathbf{0}. \quad (18)$$

The vector $\mathbf{e}_3 = \mathbf{e}_1 - \mathbf{e}_2$ is a noise-vector with at most $\alpha_t$ erasures and $2\beta_t$ errors. Therefore, there are $\alpha_t + 2\beta_t$ noisy edges and it is detectable due to Theorem 2, even if the receiver $t$ does not know the erasure-pattern. Also, $\mathbf{K}_t \mathbf{e}_3 \neq \mathbf{0}$. The equation (18) implies that $\mathbf{K}_t \cdot \mathbf{e}_3$ is a codeword. Thus, for each $\mathbf{G}_t \mathbf{u}_1 \in S_t^C$, there is a vector $\mathbf{G}_t \mathbf{u}_2 \in S_t^C$, $\mathbf{u}_1 \neq \mathbf{u}_2$, such that, $\mathbf{G}_t \mathbf{u}_1 + \mathbf{K}_t \mathbf{e}_3 = \mathbf{G}_t \mathbf{u}_2$, implying that $\mathbf{e}_3$ is not detectable, which is a contradiction. □

For the syndrome-based decoding in presence of mixture of error and erasure, the receiver $t$ constructs a look-up table for each erasure-pattern, $\Phi_{ers}$, of size $\alpha_t \leq \delta_t$. This look-up table consists of all correctable coded noise-vectors with erasure-pattern $\Phi_{ers}$. Also, the corresponding syndrome for each coded noise-vector is stored in the table. Briefly, the decoding process at receiver $t$ for a received vector $\mathbf{z}_t$ is as follows:

1- If the size of erasure-pattern, $\alpha_t$, is equal to or less than $\delta_t$, compute the syndrome. Otherwise, the received vector is not correctable by bounded distance decoding.

2- Locate the syndrome in the look-up table of the erasure-pattern. If the syndrome is found, the corresponding encoded noise-vector, $\mathbf{c}_t$, is obtained. Otherwise, the received vector is not correctable by bounded distance decoding.

3- Decode the input data: $\overline{\mathbf{u}} = \mathbf{G}_t^{-1}(\mathbf{z}_t - \mathbf{c}_t)$.

As we shall demonstrate in section VII, the proposed syndrome-based decoding leads to a substantially smaller complexity as compared to an exhaustive search scheme. In the next section, we will present a three-stage scheme which further enhances the decoding efficiency.

## V. THREE-STAGE SYNDROME-BASED DECODING

In this section, we design an efficient three-stage syndrome-based network decoding algorithm. The three stages are: noise detection, finding noise-pattern, and finding noise-vector. This eliminates the need for a look-up table, and computes the noise-vector with manageable complexity. Here, we consider only error, and no erasure, for simplicity. Note that, in the same direction, a decoding scheme for the mixture of errors and erasures may be obtained.

*A) Definitions*

We define $\mathbf{D}_t = \mathbf{H}_t^\tau \mathbf{K}_t = [\mathbf{d}_1, \cdots, \mathbf{d}_{|E^a|}]$, a $\delta_t \times |E^a|$ matrix. Considering the error-pattern $\Phi = \{j_1, \ldots, j_b\}$, $\mathbf{D}_t^\Phi = [\mathbf{d}_{j_1}, \ldots, \mathbf{d}_{j_b}]$ is a matrix of the columns of the matrix $\mathbf{D}_t$ corresponding to the edges identified by $\Phi$. The error-space of the error-pattern $\Phi$ is defined as $S(\Phi) = span(\mathbf{d}_{j_1}, \ldots, \mathbf{d}_{j_b})$, which is a subspace of $F_q^{h_t - k}$. Correspondingly, the dual space of $S(\Phi)$, with a dimension $\hat{b}$, is denoted by $\hat{S}(\Phi)$, that is spanned by $\hat{b}$ independent vectors: $\hat{\mathbf{d}}_1, \ldots, \hat{\mathbf{d}}_{\hat{b}}$. We construct the parity check matrix of the error-pattern as $\hat{\mathbf{D}}_t^\Phi = [\hat{\mathbf{d}}_1, \ldots, \hat{\mathbf{d}}_{\hat{b}}]$.

*B) Three-stage Syndrome-based Decoding*
The three stages of the scheme are as follows:

1- Error detection: In this stage, the syndrome is computed. If the syndrome is zero, the decoder accepts the received vector as correct, otherwise, it goes to the next stage.

2- Finding the error-pattern: In this stage, we employ the parity check matrices to find error-patterns by checking the syndrome vector as follows:

$$I(\mathbf{s}_t \in S(\Phi)) = I\{(\hat{\mathbf{D}}_t^{\Phi})^{\tau} \mathbf{s}_t = \mathbf{0}\}, \quad (19)$$

where $I(p)$ is the indicator function equal to 1 or 0, if $p$ is true or false, respectively. Using Equation (19), we check the syndrome for all error-pattern candidates of size $\beta_t = \lfloor \delta_t / 2 \rfloor$. If there is a candidate $\Phi$, for which the indicator function is equal to one, it is considered as the error-pattern and we go to the third stage. Otherwise, the error-vector is not correctable by the bounded-distance decoder.

3- Finding error values: In this stage, we find the error-vector, due to the identified error-pattern, $\Phi$. As the positions of the errors are now known, the syndrome vector $\mathbf{s}_t = \mathbf{D}_t \mathbf{e}$ can be considered as follows,

$$\mathbf{D}_t^{\Phi} \cdot \mathbf{e}^{\Phi} = \mathbf{s}_t \quad (20)$$

which indicates a set of $\delta_t$ (size of syndrome vector) equations. Also, the number of unknown variables is equal to the size of the error-pattern, $\beta_t = \lfloor \delta_t / 2 \rfloor$. Thus, these equations are enough to find the error values. Therefore, the vector $\mathbf{e}^{\Phi}$ is computed at receiver $t$. Obviously, in general, some error values may be zero.

Note that, there may be more than one error-pattern of size $\beta_t = \lfloor \delta_t / 2 \rfloor$, for which the indicator function is one. In this case, these error-patterns are equivalent, *i.e.*, the consequent error-vectors, computed from Equation (20), result the same coded error-vectors. This issue is elaborated in the next section.

In [23][24], with the same objective of reducing the complexity of network decoding compared to an exhaustive search, Gaussian elimination is employed to "somewhat separate the error message part from the source message part". They present a fast decoding scheme that for certain error events can locate the error positions first to simplify error computation. The scheme is packet-based and assumes that (i) the error-pattern remains unchanged over a packet, i.e., when a packet is transmitted, errors may occur only in certain network edges and, (ii) a sufficient number of errors in these positions occur within a packet. Specifically, the receiver can correct errors in $\omega \leq \lfloor \delta_t / 2 \rfloor$ edges within a packet, if it receives $\omega$ independent error-vectors in $\omega$ symbols within a packet in $\omega$ edges. Naturally, such assumptions constrain the performance. Notably, due to the second assumption, the scheme does not attain the packet-based refined Singleton bound. However, in the proposed three-stage scheme, the receiver can find the error-pattern and correct the received vector on a symbol-by-symbol basis. Based on Theorem 6, the error correction is guaranteed, as long as the number of edge errors at a certain symbol duration is at most $\lfloor \delta_t / 2 \rfloor$ (attaining the refined Singleton bound), disregarding the number and pattern of errors in other symbol durations within a packet.

In [20], a decoding scheme is presented at the packet level to achieve the packet-based Singleton bound, as the error pattern is assumed constant over a packet. The decoding delay grows with the number of symbols in a packet. For decoding a packet of size $K$ (symbols), the complexity is $O(K^3 h^3)$. The scheme does not guarantee achieving the bound, but it approaches the bound with high probability for sufficiently large packet size and symbol field size. The scheme sends additional symbols to inform the receivers of the transfer matrices of the network, but this is not a significant overhead, as the packet size is great.

## VI. COMPLETE ML DECODING

*A) Definitions*

Consider two error-patterns $\Phi = \{j_1, ..., j_b\}$ and $\Phi' = \{j'_1, ..., j'_{b'}\}$ within the network. The two error-patterns are considered to be equivalent if $span(\mathbf{k}_{j_1}, ..., \mathbf{k}_{j_b}) = span(\mathbf{k}_{j'_1}, ..., \mathbf{k}_{j'_{b'}})$, in which $\mathbf{k}_j$ is the $j$'th column of the matrix $\mathbf{K}_t$ (A similar case is considered in Definition 3 of [23]). Also, we refer to two error-vectors as equivalent, if they correspond to the same coded error-vector. For a coded noise-vector $\mathbf{c}_t$, we define *error-vector-set*, $\Omega_t(\mathbf{c}_t)$, as the set of its corresponding noise-vectors: $\Omega_t(\mathbf{c}_t) = \{\mathbf{e} \in F_q^{|E^a|} \mid \mathbf{K}_t \mathbf{e} = \mathbf{c}_t\}$. Also, a set of equal size vectors is said to be $k$-independent, if any subset of $k$ or less vectors are linearly independent [12].

*B) Motivations*

The BNEC code is designed to attain the bounded-distance decoding capabilities, in which a determined maximum number of errors and erasures in the network edges can be corrected. There are two main issues motivating the design of a more efficient decoder:

1-There are a large number of errors beyond the refined Singleton bound, which can be corrected. This point is studied in section VI.C and is the key to what we refer to as *complete* decoding.

2-In section VI.D, we present a BNEC *maximum likelihood* (*ML*) decoder, which is set up based on the following two ideas:

a) Unlike the classic channel coding, in which the probability of bit error in different positions are assumed the same, in a network setting, the probability of error in different network edges may be different. A BNEC ML decoder is to consider this matter to find the most probable coded error vector for a given received vector.

b) Unlike the classic channel coding, for a receiver, there may be equivalent error-patterns within the network. In other words, a BNEC ML decoder is to identify the most probable *coded* error-vector, which may be induced by a number of equivalent error-vectors.

Here, we consider only error, and no erasure, for simplicity. Note that, in the same direction, a decoding scheme for the mixture of errors and erasures can be obtained.

*C) Complete Decoding*

The number of encoded error-vectors, at the receiver $t$, with $r$ errors in the set $E_t$ is as follows,

$$N_{t,error}(r) \leq \sum_{i=1}^{r}\binom{|E_t|}{i}(q-1)^i < \binom{|E_t|}{r}.q^r, \quad (21)$$

The first inequality is due to the equivalency of some error-vectors. We can see that the number of encoded error-vectors corrected by the bounded-distance decoder, $N_{t,error}(\lfloor \delta_t/2 \rfloor)$ is much less than the number of syndromes, $q^{\delta_t}$ ($q > \binom{|E_t|}{\lfloor \delta_t/2 \rfloor}$ as we will show in Theorem 9). Even the number of encoded error-vectors at the receiver $t$, with $\delta_t - 1$ errors in the set $E_t$, $N_{t,error}(\delta_t - 1)$, is less than the number of syndrome vectors at the receiver $t$, as $q > \binom{|E_t|}{\delta_t - 1}$. Here, we study the error correction capability of the BNEC code to correct error-vectors beyond the Singleton bound.

The complete decoder for BNEC is defined, for which an error-vector with the hamming-weight $r$ is correctable at receiver $t$, if there is not any other error-vector, with a hamming-weight equal to or less than $r$, which corresponds to the same syndrome, but a different encoded error-vector at this receiver. Obviously, all of the error-vectors with hamming-weights equal to or less than $\beta_t = \lfloor \delta_t/2 \rfloor$ are correctable by the complete decoder. In the next two theorems, we assess the error correction capability of the BNEC complete decoding.

Note that, the network may have some equivalent error-pattern sets. Thus, we can select only one of these sets, with minimum cardinality, during the decoding process. Therefore, we can assume there are not any equivalent error-patterns in the consequent network, and also the set of $\mathbf{d}_i$'s are $\delta_t$-independent.

***Theorem 7:*** For an error-pattern $\Phi$ of size $b$, $\lfloor \delta_t/2 \rfloor < b \leq \delta_t - 1$, at most $\binom{|E^a|}{b}q^{2b-\delta_t}$ corresponding error-vectors with a hamming-weight of $b$, are not correctable.

*Proof-* Assume $\Phi = \{j_1,...,j_b\}$ as an error-pattern of size $b$. There are $\binom{|E^a|}{b} - 1$ other error-patterns of size $b$ in the network. We select one of them as $\Phi' = \{j'_1,...,j'_b\}$. Consider the following equation,

$$\mathbf{s}_t(\mathbf{e}^\Phi) = \mathbf{s}_t(\mathbf{e}^{\Phi'}),$$

or

$$\mathbf{D}_t^\Phi.\mathbf{e}^\Phi = \mathbf{D}_t^{\Phi'}.\mathbf{e}^{\Phi'}. \quad (22)$$

We wish to find the number of solutions to this equation, for which all the elements in the vector $\mathbf{e}^\Phi$ are non-zero. The equation can be written as follows,

$$\sum_{i=1}^{b}\mathbf{d}_{j_i}e_i^\Phi = \sum_{i=1}^{b}\mathbf{d}_{j'_i}e_i^{\Phi'}. \quad (23)$$

Assume $r$ edges are the same in the two error-patterns. We have,

$$\sum_{i=1}^{b-r}\mathbf{d}_{j_i}e_i^\Phi + \sum_{i=b-r+1}^{b}\mathbf{d}_{j_i}\overline{e}_i - \sum_{i=1}^{b-r}\mathbf{d}_{j'_i}e_i^{\Phi'} = \mathbf{0}, \quad (24)$$

in which we assume $j_i = j'_i, i = b - r + 1,...,b$, without loss of generality, and also $\overline{e}_i = e_i^\Phi - e_i^{\Phi'}$. Based on Lemma 1 in the Appendix, the number of solutions to Equation (24), $\eta$, is given by

$$\eta \begin{cases} = 1 & 2b - r \leq \delta_t \\ \leq q^{2b-r-\delta_t} & 2b - r > \delta_t \end{cases} \quad (25)$$

In this equation, the only solution for the case of $2b - r \leq \delta_t$, is all zero. Thus, the number of solutions to Equation (24) for which the elements $e_i^\Phi$ are non-zero is given by,

$$\eta' \begin{cases} = 0 & 2b - r \leq \delta_t \\ \leq q^{2b-r-\delta_t} & 2b - r > \delta_t \end{cases} \quad (26)$$

Now, we decompose the second term in the left side of equation (24) as

$$\sum_{i=b-r+1}^{b}\mathbf{d}_{j_i}\overline{e}_i = \sum_{i=b-r+1}^{b}\mathbf{d}_{j_i}e_i^\Phi - \sum_{i=b-r+1}^{b}\mathbf{d}_{j_i}e_i^{\Phi'}.$$

For fixed $\overline{e}_i$'s, the equation has at most $q^r$ solutions for $e_i^\Phi$'s and $e_i^{\Phi'}$'s. Thus, the number of solutions is at most $\eta'' = \eta'.q^r$,

$$\eta'' \begin{cases} = 0 & 2b - r \leq \delta_t \\ \leq q^{2b-\delta_t} & 2b - r > \delta_t \end{cases}. \quad (27)$$

Considering all error-patterns of size $b$, the result of the theorem is obtained. □

The number of error-vectors with the hamming-weight $b$, corresponding to an error-pattern of size $b$, is equal to $(q-1)^b < q^b$. Considering all error-patterns, we can find the probability of correction at the receiver $t$, in the presence of error-vectors with the hamming-weight of at most $b$, $P_{c,t}(b)$, as follows,

$$P_{c,t}(b) \begin{cases} = 1 & 0 \leq b \leq \lfloor \delta_t/2 \rfloor \\ \geq 1 - \binom{|E^a|}{b}q^{b-\delta_t} & \lfloor \delta_t/2 \rfloor < b \leq \delta_t - 1 \end{cases} \quad (28)$$

Note that in [23], with the same objective of further error correction, the authors present the decoding capabilities of their packet-based decoding scheme with an expression similar to (28). As we can see in equation (28), $P_{c,t}(\delta_t - 1) \geq 1 - \frac{1}{q}\binom{|E^a|}{b}$, which is sufficiently close to 1 for a sufficiently large field size. We can obtain the following interesting theorem.

***Theorem 8:*** The complete decoder for BNEC code, can correct error-vectors with at most $\delta_t - 1$ additive random errors, with probability sufficiently close to 1, for a sufficiently large field size. □

Note that, doubling-up the field size, is equivalent to adding one bit to the symbols. When transmission is at the bit level, this results in a small increase in the probability of symbol error (linear increase by the number of bits per symbol). However, as evident from Equation (28) for $b = \delta_t - 1$, the probability of unsuccessful decoding is reduced by a factor of 2. In other words, the probability of symbol error is increased as a function of $\log_2 q$, whereas, the probability of unsuccessful decoding decreases linearly with $q$.

*D) Complete ML Decoding*
In this section, we design a complete ML decoding scheme for BNEC, based on syndrome. The likelihood function at the decoder of the receiver $t$ is $L(\mathbf{v}_t) = p_{\mathbf{Z}_t|\mathbf{V}_t}(\mathbf{z}_t | \mathbf{v}_t)$, and the ML decoder is given by,

$$\bar{\mathbf{v}}_t = \arg\max \{p_{\mathbf{Z}_t|\mathbf{V}_t}(\mathbf{z}_t | \mathbf{v}_t)\}. \qquad (29)$$

Since, $\mathbf{z}_t = \mathbf{v}_t + \mathbf{c}_t$, the likelihood function can be considered as $L(\mathbf{v}_t) = \Pr\{\mathbf{c}_t = \mathbf{z}_t - \mathbf{v}_t\}$. Thus, it is sufficient to determine the probability of the encoded error-vector due to the received vector and a given $\mathbf{v}_t$ (motivation 2-*b*). The probability of an encoded error-vector, $\mathbf{c}_t$, is the summation of the probabilities of the error-vectors of the set $\Omega_t(\mathbf{c}_t)$:

$$\Pr(\mathbf{c}_t) = \sum_{\mathbf{e} \in \Omega_t(\mathbf{c}_t)} \Pr(\mathbf{e}), \qquad (30)$$

where the probability of the error-vector $\mathbf{e}$, with $b$ errors in the edges $l_1, \ldots, l_b$ is as follows (motivation 2-*a*),

$$\Pr(\mathbf{e}) = \frac{1}{(q-1)^b} \prod_{i=1}^{b} p_{err}(l_i), \qquad (31)$$

based on the assumed error model in section II. Here, we employ the syndrome to find the most probable encoded error-vector, in the following two steps:
1- Identify the possible encoded error-vectors corresponding to each computed syndrome.
2- Determine the most probable encoded error-vector candidate.

*Basic Syndrome-based Complete ML Decoding*
For this scheme, we find the most probable encoded error-vector, up to a hamming-weight of $\delta_t$, for each received syndrome and store it in a look-up table. Next, the input vector is obtained as $\bar{\mathbf{u}} = \mathbf{G}_t^{-1}(\mathbf{z}_t - \mathbf{c}_t)$.

*Three-stage Syndrome-based Complete ML Decoding*

In this case, at the receiver $t$, we check the error-patterns of size $\delta_t - 1$ within the active edge set of the receiver, to find the patterns that set the indicator function in Equation (19). If a candidate is found, we compute the related error-vector, and encoded error-vector. We set aside the edges that correspond to non-zero elements in the error-vector within the active edge set of the receiver $t$, and continue to check the other error-patterns in the updated edge set. At the end, we select the most likely encoded error-vector candidate. In case, there is not any error-pattern candidate of size $\delta_t - 1$, we can select the most likely error-pattern of size $\delta_t$.

Note that, if the field size $q$ is a large number, for the presented error model, due to Equations (30) and (31), the complete ML decoder may consider only the error-vectors with a small or minimum hamming-weight in $\Omega_t(\mathbf{c}_t)$ to compute $\Pr(\mathbf{c}_t)$ with a reasonable approximation.

Recently in [29], ML decoding is suggested with a focus on two-user binary unicast multiple access relay channel with network coding at the relay [30]. Specifically, the authors assess the effect of the availability of state information for channels within the network (not directly connected to the receiver) on the performance of the ML decoder. The decoding scheme at the receiver aims at exploiting the equivalent *error-vectors* in the network (motivation 2-*b* in section VI.B). Immediate application of this scheme in more general networks (larger network and field sizes) results in a decoding complexity that grows exponentially with the number of network edges.

VII. PERFORMANCE AND COMPLEXITY ANALYSIS

*A) Performance Analysis*
*A-1) Error Detection Performance:* The probability of error detection at receiver $t$ for BNEC is as follows,

$$P_{d,t} \geq \sum_{\alpha_t \leq \delta_t} \Pr(\mathbf{e}) \qquad (32)$$

Consider a uniform network with edge error probability, $\rho_1$, we have,

$$P_{d,t} \geq \sum_{i=0}^{\delta_t} \binom{|E_t|}{i} \rho_1^i (1-\rho_1)^{|E_t|-i} \qquad (33)$$

*A-2) Decoding Performance:* We presented three approaches for BNEC decoding: bounded-distance, complete and complete ML decoding. For BNEC bounded-distance decoding, the probability of error correction at receiver $t$ in a uniform network, with erasure probability, $\rho_2$, is as follows,

$$P_{c,t}^{BD} \geq \sum_{\alpha=0}^{\delta_t} \sum_{i=0}^{\alpha} \sum_{j=0}^{\lfloor(\delta_t-\alpha)/2\rfloor} \binom{|E_t|}{i}\binom{|E_t|-i}{j} \rho_2^i \rho_1^j (1-\rho_2)^{|E_t|-i}(1-\rho_1)^{|E_t|-j} \qquad (34)$$

If we consider only error in the network and no erasure, we have

$$P_{c,t}^{BD} \geq \sum_{i=0}^{\lfloor \delta_t/2 \rfloor} \binom{|E_t|}{i} \rho_1^i (1-\rho_1)^{|E_t|-i} \quad (35)$$

For complete decoding in a uniform network, the probability of error correction at receiver $t$ is as follows,

$$P_{c,t}^{CD} \geq P_{c,t}^{BD} + \sum_{i=\lfloor \delta_t/2 \rfloor+1}^{\delta_t-1} \left(1 - \frac{1}{q^{\delta_t-i}} \binom{|E^a|}{i}\right) \binom{|E_t|}{i} \rho_1^i (1-\rho_1)^{|E_t|-i}, \quad (36)$$

which indicates a noticeable improvement, when compared to the bounded-distance decoding. Note that, the benefits of complete ML decoding can be analyzed for different non-uniform networks.

*B) Complexity and Memory Size*

*B-1) Error Detection:* Employing exhaustive search, one can find all possible error-free output vectors for all possible input vectors and declare an error, if the received vector is not among the $q^k$ error-free outputs. Thus, the computational complexity order is $O(k.\log_2(q))$, while the memory requirement is $O(q^k)$. For the proposed syndrome-based error detection, it is enough to compute the syndrome. If the syndrome is nonzero, the receiver declares an error. Thus, the complexity corresponds to syndrome computation and no look-up table is required.

*B-2) Bounded-distance Decoding:* For a given erasure-pattern, $\Phi_{ers}$, the number of correctable coded noise-vectors by bounded-distance decoding is as follows,

$$N_{ccn}^{\Phi_{ers}}(t) \leq N_{cn}^{\Phi_{ers}}(t) = \sum_{i=0}^{\lfloor (\delta_t-|\Phi_{ers}|)/2 \rfloor} \binom{|E_t|-|\Phi_{ers}|}{i}(q-1)^i. \quad (37)$$

in which $N_{cn}^{\Phi_{ers}}(t)$ is the number of correctable noise-vectors by bounded distance decoder for the erasure-pattern $\Phi_{ers}$ at receiver $t$. Also, the number of all correctable encoded noise-vectors is

$$N_{ccn}(t) \leq N_{cn}(t) = \sum_{j=0}^{\delta_t} \sum_{i=0}^{\lfloor (\delta_t-j)/2 \rfloor} \binom{|E_t|}{j}\binom{|E_t|-j}{i}(q-1)^i. \quad (38)$$

The number of correctly decodable received vectors for erasure-pattern $\Phi_{ers}$ is

$$N_{rec}^{\Phi_{ers}}(t) = q^k \cdot N_{ccn}^{\Phi_{ers}}(t). \quad (39)$$

The number of all correctly decodable received vectors is

$$N_{rec}(t) = q^k \cdot N_{ccn}(t). \quad (40)$$

Considering Theorem 6, the number of syndromes at receiver $t$ for erasure-pattern $\Phi_{ers}$ is

$$N_s^{\Phi_{ers}}(t) = N_{ccn}^{\Phi_{ers}}(t). \quad (41)$$

Employing exhaustive search, the receiver constructs a look-up table of size $N_{rec}^{\Phi_{ers}}(t)$ for each erasure-pattern $\Phi_{ers}$. The total memory size is $N_{rec}(t)$. The complexity of decoding for this erasure-pattern is $O(\log_2 N_{rec}^{\Phi_{ers}}(t)) = O(k.\log_2 q + \log_2 N_{ccn}^{\Phi_{ers}}(t))$.

For the proposed syndrome-based decoding, the look-up table size for each erasure-pattern $\Phi_{ers}$, is reduced to $N_s^{\Phi_{ers}}(t)$ leading to a total memory size of $N_{cn}(t)$. The complexity, in this case, is $O(\log N_{cn}^{\Phi_{ers}}(t))$. In presence of only errors, the memory size for basic syndrome-based decoding is $O\left(q^{\lfloor \delta_t/2 \rfloor}\binom{|E|}{\lfloor \delta_t/2 \rfloor}h_t \log q\right)$ bits, and also, the complexity is $O\left(\lfloor \delta_t/2 \rfloor \log_2 q + \log_2 \binom{|E_t|}{\lfloor \delta_t/2 \rfloor}\right)$. It is evident that the memory size in the proposed syndrome-based BNEC decoding is a factor of $q^k$ smaller, when compared to the exhaustive search, but still grows exponentially with the redundancy order, $\delta_t$.

For the three-stage syndrome-based BNEC decoding, considering only error and no erasure, we store the matrix $\hat{\mathbf{D}}_t^{\Phi}$, of size at most $\delta_t \times \lceil \delta_t/2 \rceil$ at receiver $t$, for all error-patterns of size $\lfloor \delta_t/2 \rfloor$. Thus, the memory size is $\delta_t \lceil \delta_t/2 \rceil \cdot \binom{|E_t|}{\lfloor \delta_t/2 \rfloor}$ symbols or $\delta_t \lceil \delta_t/2 \rceil \cdot \binom{|E_t|}{\lfloor \delta_t/2 \rfloor} \cdot \log q$ bits. Thus, the memory size of the three-stage scheme is substantially smaller than that of the basic syndrome-based scheme.

The complexity of the three-stage scheme corresponds to the complexity of the second stage, i.e., finding the error-pattern. The indicator function of an error-pattern is computed based on the multiplication of a matrix of size $\lceil \delta_t/2 \rceil \times \delta_t$ by a vector of size $\delta_t \times 1$, due to equation (19). Thus, the complexity is $O\left(\lceil \delta_t/2 \rceil \delta_t^2 \binom{|E_t|}{\lfloor \delta_t/2 \rfloor}\right)$, in $F_q$. Compared with the basic syndrome-based BNEC decoding scheme, the proposed three-stage decoding scheme, substantially reduces the memory requirement, at the cost of a modest increase of the computational complexity.

*B-3) Complete ML Decoding:* We consider the proposed basic and three-stage syndrome-based schemes for complete ML decoding. The basic scheme identifies a coded error-pattern of size $h_t \times 1$ for each of the syndromes. Thus, the look-up table consists of $q^{\delta_t}$ coded noise-vectors, and memory requirement is $O(q^{\delta_t} h_t \log q)$ bits. The error-vectors may be stored in the order of the corresponding syndromes. There is a noise-vector for each syndrome, and it is enough to consider the syndrome as the relative address of the position of the error-vector in the memory. Thus, the complexity is negligible.

For three-stage complete ML decoding, we store the matrix $\hat{\mathbf{D}}_t^{\Phi}$ for all error-patterns of size $\delta_t - 1$. The size of each matrix is $\delta_t \times 1$. Thus, the total memory requirement is $\delta_t \cdot \binom{|E_t|}{\delta_t - 1} \cdot \log_2 q$ or $O(\delta_t \cdot q \cdot \log_2 q)$, which is much less than that of the basic scheme.

In three-stage complete ML decoding, the indicator function of a noise-pattern is computed based on the multiplication of a matrix of size $1 \times \delta_t$ by a vector of size $\delta_t \times 1$. Thus, the complexity is $O\left(\delta_t^2 \binom{|E_t|}{\delta_t - 1}\right)$, in $F_q$; Although, this is greater than that of the basic scheme, considering the savings in memory size, the proposed three-stage decoding scheme provides an efficient solution for BNEC decoding.

## VIII. A SCHEME TO DESIGN BNEC

In this section, based on the BNEC design strategy presented in section III, to compute the desired matrices and design the BNEC code, we present an efficient algorithm based on the preservative approach of [5]. Note that, other approaches such as algebraic approach of [3] employed in [18], and the random approach of [4] can also be employed for the design.

*Design Procedure*
1) Initialization: We add a virtual node, $s'$, to the network, and also add $h_{max}=\max(h_t)$ virtual directed edges from the node $s'$ to the source node $s$. These edges are noise-free with transmission rate one. Assume that node $s'$ generates the input vector **u** of size $k \times 1$. The new network has $h_{max} + |E^a|$ edges, numbered as described in section II (the virtual edges are numbered from 1 to $h_{max}$).

2) Finding paths: The first stage of the code design is to find $h_t$ paths with no common edges, referred to as edge-disjoint paths, from the node $s'$ to each receiver $t \in T$. Note that, the paths of two receivers may join in some edges. Without loss of generality, we assume the first edges for $h_t$ paths of receiver $t$ are $l_1, ..., l_{h_t}$.

For each edge $l$, the set of all receivers that employ $l$ in one of the paths is denoted by $T(l)$. $Prev(l,t)$ is the previous edge of the edge $l$ in the path to $t$, passing through $l$.

3) Designing encoding vectors: In this stage the local and global encoding vectors of the active edges are designed. We begin with several definitions. We define the set $C_{t,i}$, composed of the edges in the $h_t$ edge-disjoint paths transmitting symbols for the receiver $t$, considered in step $i$ in which the local encoding vector of the edge $l_{i+h_{max}}$ is designed. The set is initialized by the first $h_t$ virtual edges, $l_1,...,l_{h_t}$, for receiver $t$, denoted by $C_{t,0}$. Also, we define the set $\mathbf{B}_{t,i} = [\mathbf{G}_{t,i}, \mathbf{K}_{t,i}]$, in step $i$, as a matrix whose rows are the global encoding vectors of the edges in $C_{t,i}$, and $\mathbf{B}_{t,i}^\Phi = [\mathbf{G}_{t,i}, \mathbf{K}_{t,i}^\Phi]$ as the shortened version for the erasure-pattern $\Phi$. Thus, the matrix $\mathbf{B}_{t,0}$ is initialized by the global encoding vectors of $l_1,...,l_{h_t}$. These edges are noise-free, thus $\mathbf{K}_{t,0} = \mathbf{0}$. Also, the matrix $\mathbf{G}_{t,0}$ is initialized by $h_t$ rows, constructing a $k$-independent vector set.

At step $i$, $i=1,...,|E^a|$, the set $C_{t,i}$ is constructed for all receivers. The edge $l_i$ replaces $Prev(l_i,t)$ for all $t$ employing the edge $l_i$. For other receivers the set remains unchanged: $C_{t,i} = C_{t,i-1}$. The local encoding vector for the edge $i$, $\mathbf{m}_i$ is selected randomly. The vector $\mathbf{gev}_i$ is computed from Equation (3) and replaces $\mathbf{gev}_{n(Prev(l_i,t))}$. The matrices $\mathbf{G}_{t,i}$ and $\mathbf{K}_{t,i}$ are updated with the new vectors, and thus, a candidate for $\mathbf{B}_{t,i} = [\mathbf{G}_{t,i}, \mathbf{K}_{t,i}]$ is determined for all $t$. For each $t$ and $\Phi$, if the rank of the matrix $\mathbf{K}_{t,i}^\Phi$ is denoted by $p_t(\Phi,i)$, the matrix $\mathbf{B}_{t,i}^\Phi = [\mathbf{G}_{t,i}, \mathbf{K}_{t,i}^\Phi]$ is verified for $(k + p_t(\Phi,i))$-independence for the set of rows. Note that, there is no need to decode errors, thus there is no problem if $p_t(\Phi,i) < \delta_t$ at the end of the scheme at certain receivers for some $\Phi$. Finally, at the end of the scheme, all the local encoding vectors are designed, and we have $[\mathbf{G}_t, \mathbf{K}_t] = B_{t,|E^a|}$. The pseudo-code of the proposed scheme is shown in Figure 3.

---

*Initialization*------------------------------------------------------
**for each** $t \in T$ **do**
    initialize $C_{t,0}$ and $\mathbf{G}_{t,0}$
    $\mathbf{K}_{t,0}=\mathbf{0}$
**for each** erasure-pattern $\Phi$ of size $\delta_t$ **do**
    $p_t(\Phi,0)=0$
    $\mathbf{K}_{t,i}^\Phi = \mathbf{0}$

*Designing Local Encoding Vectors*-------------------------
**for** $i=1:|E^a|$ **do**
    **for each** $t \in T(l_i)$
        $C_{t,i} \leftarrow C_{t,i} \setminus \{Prev(l_i,t)\} \cup \{l_i\}$

    *Random Design Loop*---------------------------------
    **select** $\mathbf{m}_i$ randomly
    $\mathbf{gev}_i = \sum_{l \in In(l_i)} m_i(l)\mathbf{gev}_{n(l)} + e_i \cdot \mathbf{1}_{k+|E^a|,k+i}$
    update $\mathbf{G}_{t,i}$, $\mathbf{K}_{t,i}$
    **for each** erasure-pattern $\Phi$ **do**
        **if** $[\mathbf{G}_{t,i}|\ \mathbf{K}_{t,i}^\Phi]$ is not $(k+p_t(\Phi,i))$-independent
            **go to select**
    *End of Random Design Loop*-----------------------------

Figure 3- The pseudo-code of the proposed scheme for BNEC($h+,k$)

The field size can be set using the following theorem.

***Theorem 9***: A valid local encoding vector for each edge exists in BNEC($h+,k$), if the size, $q$, of the employed finite field $F_q$ satisfies the following condition

$$q \geq \sum_{t \in T} \binom{|E^a|}{\delta_t} \cdot \binom{h_t}{k}. \tag{42}$$

*Proof-* The local encoding vector of the edge $l_i$ is verified for all receivers in $T(l_i)$ and all erasure-patterns with a hamming-weight of $\delta_t$ (note that $|T| \geq |T(l)|$). The number of independence tests in ($k+p_t(\Phi,i)$)-independence verifications in an edge $l_i$ to check the local encoding vector is at most $\binom{h_t}{k}$. Thus, the number of independence tests in an edge is at most $\sum_{t \in T} \binom{|E^a|}{\delta_t} \cdot \binom{h_t}{k}$. The field size is to be equal or greater than the number of independence tests for the design of the local encoding vector of a network edge [13], which leads to the Equation (42). □

For each edge, we select the local encoding vector randomly, and verify independence for all cases. The complexity of this scheme is due to this part as quantified below.

***Theorem 10***: The expected number of operations, in the field $F_q$, for designing BNEC($h+,k$) code is $O(|E^a| \cdot \binom{|E^a|}{\delta_{max}} \cdot |T| \cdot h_{max}^3)$, where $\delta_{max} = \max_{t \in T}(\delta_t)$.

*Proof-* The algorithm finds the local encoding vector for all active edges. In each active edge, the local encoding vector is verified for all erasure-patterns with a hamming-weight of $\delta_t$, which leads to the factor $\binom{|E^a|}{\delta_{max}}$. Also, verification is done for all receivers in the set $T(l_i)$ for the edge $l_i$, which leads to the factor $|T| \geq |T(l_i)|$. For each verification, ($k+p_t(\Phi,i)$)-independence is checked, which is at most as complex as the computation of the determinant of an $h_{max} \times h_{max}$ matrix, which is equal to $O(h_{max}^3)$ ($k+p_t(\Phi,i) \leq h_{max}$). □

In [21], a network error correcting code is presented which also employs the approach of [5] for code design. In the algorithm of [21], the set of paths within the network is to be found for each erasure-pattern, which increases the design complexity. Here, one set of paths are identified instead for all erasure-patterns. Another design algorithm for a network error correcting code to achieve the refined Singleton bound is presented in [18], based on the algebraic approach of [3].

## IX. CONCLUDING REMARKS

In this paper, we present a block network error control coding framework, BNEC($h+,k$), for multicast in a directed acyclic network with imperfect links. The ability to correct and detect a mixture of symbol errors and packet erasures on the network edges are studied. In the presented setting, the error-pattern may vary within a packet duration, which is necessary for achieving the refined Singleton bound. This is in contrast to the packet-based refined Singleton bound, characterized when the error-pattern is assumed fixed during a packet interval. A design algorithm is presented for a BNEC code to attain the bounded-distance capabilities, based on the refined Singleton bound.

The idea of syndrome-based decoding for block network error correcting codes is presented. By removing the effect of the input vector, the memory requirement for decoding shrinks by a factor of $q^k$. An efficient three-stage syndrome-based decoding scheme is also introduced, in which prior to finding the error values, the position of the edge errors are identified based on the error spaces at the receivers. The scheme eliminates the need for a look-up table, at the cost of a modest increase in the computational complexity. The great gap between the number of errors corrected by the bounded-distance decoder and the number of syndromes, lead us to devise a complete syndrome-based decoder, which results in a new bound for error correction using a BNEC code. We demonstrate that for a random additive error, a code with redundancy order $\delta_t$ at the receiver $t$, can correct $\delta_t$-1 errors with a probability sufficiently close to 1, if the field size is sufficiently large. Finally, the complete ML decoding is studied for BNEC. Unlike classic channel coding, in which all the bits usually experience the same error probabilities, in a network setting, the probabilities of error in different edges may be different. Thus, the number of edge errors is not a sufficient statistic for ML decoding. The proposed BNEC ML decoder exploits this fact and the equivalency of certain edge errors within the network at a particular receiver.

## APPENDIX

***Lemma* 1**: Consider a $\delta_t$-independent set of vectors, $\{\mathbf{d}_i\}_{i=1}^m$ of size $\delta_t \times 1$. The number of solutions, $\eta$, for $e_i$'s in the equation $\sum_{i=1}^m \mathbf{d}_i e_i = \mathbf{0}$, is as follows:

$$\eta \begin{cases} =1 & m \leq \delta_t \\ \leq q^{m-\delta_t} & m > \delta_t \end{cases}.$$

*Proof-* For $m \leq \delta_t$, all the coefficients are to be zero, and there is only one solution. For $m > \delta_t$, the equation can be written as follows,

$$\sum_{i=1}^{\delta_t} \mathbf{d}_i e_i = -\sum_{i=\delta_t+1}^m \mathbf{d}_i e_i$$

The RHS has $m - \delta_t$ scalar variables, and has $q^{m-\delta_t}$ possible values. We show that for each value of the RHS, there is only one solution for the LHS. Assume there are two solutions for the left side,

$$\sum_{i=1}^{\delta_t} \mathbf{d}_i e_i = - \sum_{i=\delta_t+1}^{m} \mathbf{d}_i e_i$$

$$\sum_{i=1}^{\delta_t} \mathbf{d}_i e'_i = - \sum_{i=\delta_t+1}^{m} \mathbf{d}_i e_i$$

By subtracting the equations we have

$$\sum_{i=1}^{\delta_t} \mathbf{d}_i (e_i - e'_i) = \mathbf{0}.$$

Since the vectors are linearly independent, we have $e_i = e'_i$ for $i = 1,...,\delta_t$, and consequently the solutions are identical. Thus, the number of solutions is at most $q^{m-\delta_t}$. □